\documentclass[3p,authoryear,times,12pt]{elsarticle}
\usepackage{color}
\usepackage{url}
\usepackage{multirow,setspace,times,amssymb,amsmath,graphicx,color,subfigure,url}
\usepackage{dsfont}
\usepackage{threeparttable}
\usepackage{rotating}
\usepackage{dcolumn}
\usepackage{listings}
\usepackage{lscape}

\begin{document}
	
\begin{frontmatter}
				
\title{New insights into price drivers of crude oil futures markets: Evidence from quantile ARDL approach}
\author[CU]{Hao-Lin Shao}

\author[SSI]{Ying-Hui Shao \corref{cor1}}
\ead{yinghuishao@126.com}	

\author[ECNU]{Yan-Hong Yang}

\cortext[cor1]{Corresponding author.}
\address[CU]{Department of Statistics, Graduate School of Arts and Sciences, Columbia University, New York, NY 10027, USA}
\address[SSI]{School of Statistics and Information, Shanghai University of International Business and Economics, Shanghai 201620, China}
\address[ECNU]{Faculty of Education, East China Normal University, Shanghai 200062, China}
			
\begin{abstract}
This paper investigates the cointegration between possible determinants of crude oil futures prices during the COVID-19 pandemic period. We perform comparative analysis of WTI and newly-launched Shanghai crude oil futures (SC) via the Autoregressive Distributed Lag (ARDL) model and Quantile Autoregressive Distributed Lag (QARDL) model. The empirical results confirm that economic policy uncertainty, stock markets, interest rates and coronavirus panic are important drivers of WTI futures prices. Our findings also suggest that the US and China's stock markets play vital roles in movements of SC futures prices. Meanwhile, CSI300 stock index has a significant positive short-run impact on SC futures prices while S\&P500 prices possess a positive nexus with SC futures prices both in long-run and short-run. Overall, these empirical evidences provide practical implications for investors and policymakers.
\end{abstract}
\begin{keyword}
Crude oil futures; Stock markets; COVID-19 pandemic; Economic policy uncertainty; Quantile ARDL
\\
{\textit{JEL classification:C5, Q4, Q43}} 
\end{keyword}
\end{frontmatter}
	
\section{Introduction}
Crude oil futures is the most traded commodity in the world. It has been increasing crucial to financial markets and global economy. 
While there are multiple types of crude oil, the primary international benchmark crude is West Texas Intermediate (WTI) which is traded continuously on the New York Mercantile Exchange (NYMEX) from 6:00 pm to 5:00 pm EST.  
It is classified as light sweet crude oil and traded in US dollars.

Understanding the determinants of the crude oil future market is of great interest. 
Previous research mainly 
examines the contributions of supply and demand fundamentals \citep{PAN2017130, MENG2019476}, macroeconomic factors \citep{ALOUI201692, MENG2019476}, financial factors \citep{KYRTSOU2016239,LU2020104721} and political factors \citep{CHEN201642, QIN2020104851} to crude oil markets.  
Among extensive studies on the price drivers of crude oil market, 
much attention has focused on developed markets, not emerging market.

While crude oil prices used to be quoted in US dollar, China has been attempting to launch a yuan-denominated oil futures and internationalize the RMB. As the largest importer and major consumer of crude oil, China has a strong desire to establish an Asian crude oil benchmark and become an oil price setter. On March 26, 2018, China launched its crude oil futures on the Shanghai International Energy Exchange (INE), which is traded in Chinese yuan. Shanghai crude oil futures contracts (SC) is China's first commodity derivative open to international investors. 
Different from WTI crude, SC futures is medium sour crude  oil   
traded on weekdays in day sessions (9:00 am–11:30 am and 1:30 pm–3:00 pm BJT) and night sessions (9:00 pm-2:30 am BJT).

With the potential for arbitrage trade with other financial markets, 
SC has a notable increase in international involvement. It has been the third most traded crude oil futures contract in the world. And  the Chinese crude oil futures contract helps to reflect dynamic changes in crude oil supply and demand of Asia-Pacific region. Hence, we conduct a comparative empirical analysis of WTI and SC to complement studies on crude oil price drivers. 
We reconsider the influence of economic policy uncertainty (EPU), stock market, and interest rate  
on crude oil futures.

Besides the above factors, crude oil futures price remains extremely fragile to extreme events. 
Since December 2019, the COVID-19 pandemic swept across the world and 
triggered investor panic. The shock caused by the coronavirus created a turmoil in crude oil futures market. 
The unprecedented and dramatic movements in crude oil market have got ample attention \citep{MENSI2020101829,ATRI2021102061}. Nevertheless,  a few studies have addressed the long-run and short-run asymmetric impacts of COVID-19 pandemic on crude oil futures markets. 
To investigate the temporary and long-lasting influences of shocks and other factors on crude oil markets, we employ the Autoregressive Distributed Lag (ARDL) model \citep{pesaran1999pooled} and the Quantile Autoregressive Distributed Lag (QARDL) model \citep{CHO2015281} during COVID-19 pandemic period.  
The methods can simultaneously measure the potential 
asymmetric effects of variables on crude oil prices.

This paper makes the following extensions to the current literature. We examine how WTI and the newly shipped SC market reacts to factors in the long and short term. First,
We find that in the short-run, stock markets of China and US are strongly and positively correlated with crude oil markets. However, in the long-run, the fluctuations in domestic stock market have limited influence on crude oil prices. Correspondingly investors and regulators should watch stock markets performance closely. Second, we find that except stock market, other price drivers of WTI futures have no effect on SC futures. This indicates the determinants 
of Chinese crude oil futures market remain to be determined. 
	
This paper is organized as follows. Section \ref{Sec: review} reviews the relevant literature on determinations of crude oil price. Section \ref{Sec: data and method} reports source of data and describes the methodology. Section \ref{Sec: result} presents the empirical results of ARDL model and QARDL model. Section \ref{Sec: conclusion} reports the conclusion.
	
\section{Literature Review}
\label{Sec: review}
	
An array of studies have examined the time-varying effects of  supply and demand fundamentals. Some studies point that the influence of supply shocks decrease over time. Supply shocks are not the primary factors that drive the movements of crude oil prices \citep{baumeister2013time}. Oil demand plays a much more important role than the oil supply shocks \citep{kilian2009not}. After 2008 financial crisis, the roles of non-fundamentals have attracted considerable attention. The financial crisis changed market participants' expectations and mechanisms of price formation. Non commercial traders begin to play a vital role in crude oil markets \citep{JOO2020100516}.  Therefore, our study do not concentrate on the direct effects of supply and demand shocks. 
	
Constructed by \citet{10.1093/qje/qjw024}, economic policy uncertainty (EPU) index through affecting participants' beliefs and macroeconomic activities to influence crude oil price movements \citep{ALOUI201692}. USEPU plays a crucial role in determining other countries' EPU and long-run WTI spot prices \citep{YANG2019219}. Compared with other countries' EPU indexes, Global EPU and USEPU are more informative in forecasting volatility of WTI prices \citep{WEI2017141}. Existing studies have not reached a consistent conclusion in the impact of EPU on crude oil markets. Some studies find USEPU is negative correlated with WTI returns \citep{ZHANG2020750}. The positive relationship between USEPU and WTI prices has also been found \citep{LEI2019341}. In addition, the relationship between EPU index and crude oil prices might vary over time.  \citet{LEI2019341} reveal that before 2008 financial crisis, USEPU has a negative impact on WTI returns.  After the financial crisis, higher USEPU is accompanied with higher WTI returns.
	
With continuous development of international financial markets, more and more financial institutions have entered commodity markets. Financial factors play indispensable roles in crude oil related topics. Plenty of studies have found strong evidence of dependence between stock markets and crude oil markets. Standard \& Poor's 500 (S\&P500) index has been widely employed in crude oil forecasting models \citep{LU2020104721}. In addition, a persistent lead-lag relationship between S\&P500 and WTI futures has been found \citep{KYRTSOU2016239}. 
In recent years, Chinese government attempts to internationalized mainland China's financial market and attract more foreign capitals.  China's stock market also has been observed a close relationship with WTI crude oil prices \citep{GUO2021105198}. Using China Securities Index 300 (CSI300) and S\&P500 as a representative financial index for China and US respectively, \citet{ZHU2021120949} find significant risk spillovers from China and US stock markets to crude oil markets. 
    
Interest rates are often used as a proxy of macroeconomic activities in crude oil studies. Compared with short-term interest rates, crude oil prices are more sensitive to variations in long-term interest rates. Furthermore, the influence of long-term interest rates is growing with time  \citep{ARORA2013546}. There are some possible explanations of the linkage between crude oil markets and interest rates. One of explanation is based on the theory of storage. Low interest rates would require more inventories and further change supply chain of products. Thus, interest rates are related with commodity prices. An alternative explanation is that interest rates can affect the demand of crude oil \citep{PREST201863}. More specifically, interest rates reflect country's monetary policy and are highly correlated with inflation and economic prospects. Low interest rates would change participants' expectations and motivate companies to increase their spending \citep{MENSI2020100836}. Another explanation is related to financialization of commodity markets. Fluctuations in interest rates make market participants change their investment portfolios and choose to make investments in crude oil \citep{ARORA2013546}.

COVID-19 pandemic has caused severe disruptions in global financial and commodity markets. A plenty of studies have examined the effects of COVID-19 on crude oil markets. During pandemic period, crude oil markets become more inefficient \citep{MENSI2020101829}. Pandemic-related news and reports spread people's panic and increase investors' anxiety. It has been proved that the number of deaths and coronavirus panic have significant and negative impacts on WTI prices \citep{ATRI2021102061}. The influence of COVID-19 pandemic on the volatility of WTI future prices has exceeded 2008 financial crisis \citep{ZHANG2021101702}. There are relatively few studies focus on the impact of COVID-19 pandemic on SC futures market. By applying high-frequency heterogeneous autoregressive model, \citet{NIU2021102173} find coronavirus news can predict volatility of SC futures prices.

Many studies have examined the derminants of WTI crude oil futures prices. However, the potential nonlinear and asymmetric relationships between explanatory variables and WTI crude oil prices are often ignored. In addition, SC futures is a newcomer of international crude oil markets. There are few studies concentrate on it. In this study, we apply the linear ARDL model and QARDL model to investigate determinants of crude oil prices during COVID-19 pandemic period. The QARDL model has been widely employed in energy economics \citep{BENKRAIEM2018169,APERGIS2021112118,GUO2021105198}. This model can simultaneously address long-run and short-run relationships between crude oil futures and explanatory variables across quantiles. The results of QARDL model are more robust than traditional linear models. 
	
\section{Data and Methodology}	
\label{Sec: data and method}
\subsection{Data}
Our sample data comprises of daily observations of WTI futures prices, SC futures prices, US economic policy uncertainty (USEPU) index, China economic policy uncertainty (CEPU) index, S\&P500 prices, CSI300 prices, US 10-year treasury bond rate (Tbill), 1 week-Shanghai Interbank Offered Rate (SHIBOR) and Panic index. We use the closing prices of each futures and stock index. For Panic index data series, each country's first non-zero date is different. Therefore, the panel data start from the earliest day that all observations are available. For US panel, the sample period ranges from January 16th, 2020, to May 26th, 2021. For China panel, the sample period ranges from January 2nd, 2020, to May 26th, 2021. As the trading time of different markets varies, we eliminate the missing data and non-match data of each panel. The US economic policy uncertainty series come from the EPU website (http://www.policyuncertainty.com). \citet{10.1093/qje/qjw024}'s CEPU index is measured at monthly frequency. Thus, we choose to use a new released CEPU index which is constructed by \citet{huang2020measuring}. It is a daily index based on China mainland newspaper and is sourced from https://economicpolicyuncertaintyinchina.weebly.com. The US 10-year Treasury Yield rate is gatherd from US department of the treasury website (https://www.\\treasury.gov). Panic index is obtained from Ravenpack (https://coronavirus.ravenpack.com). This index measures the level of news that mentions hysteria or panic with coronavirus.  Data of other variables are collected from Wind database. In addition, we transfered all data series into natural logarithmic form.

\subsection{Methodology}
To examine the asymmetric and nonlinear effects of economic policy uncertainty, stock markets, interest rates and coronavirus panic on crude oil futures markets, we employ the ARDL model and QARDL model. The QARDL model is developed by \citet{CHO2015281}, it extends the ARDL-ECM model into quantile regression context. 

The traditional linear ARDL model is written as follows:
\begin{equation}\label{eq:1}
\begin{aligned}
Oil_{t}&=
\alpha+\sum_{i=1}^{p}\phi_{i}Oil_{t-i}+\sum_{i=0}^{q_{1}}\omega_{i}EPU_{t-i}+\sum_{i=0}^{q_{2}}\lambda_{i}S\&P500_{t-i}+\sum_{i=0}^{q_{3}}\theta_{i}CSI300_{t-i}+\sum_{i=0}^{q_{4}}\psi_{i}Interest_{t-i}\\
&+\sum_{i=0}^{q_{5}}\delta_{i}Panic_{t-i}+\varepsilon_{t}
\end{aligned}
\end{equation}
where $Oil_{t}$ refers to WTI futures prices and SC futures prices, $EPU_{t-i}$ refers to USEPU and CEPU, and $Interest_{t-i}$ denotes Tbill and SHIBOR, respectively. $\varepsilon_{ t}$ is the error term which is defined as $Oil_{t}- E[{Oil_{t}}/F_{t-1})]$. $F_{t-1} $ is the smallest $\sigma$-field formed by \{$EPU_{t}$, $S\&P500_{t}$, $CSI300_{t}$, $Interest_{t}$, $Panic_{t}$, $Oil_{t-1}$, $EPU_{t-1}$, $S\&P500_{t-1}$, $CSI300_{t-1}$, $Interest_{t-1}$, $Panic_{t-1}$,...\}. $p$, $ q_{1}$, $q_{2}$, $q_{3}$, $q_{4}$, $q_{5}$  are the orders of lag. The ARDL model has some advantages over other normal linear model. It allows mixed orders of integration ($I_{0}$ and $I_{1}$) and the lag lengths of variables can be different. In addition, with appropriate lags, the ARDL model can correct the problem of endogeneity and serial correlation \citep{pesaran1999pooled}. However, the linear ARDL model can only examine the average effects among variables. Therefore, following \citet{CHO2015281}, the linear ARDL model is extended into quantile regression context as following equation:
\begin{equation}\label{eq:2}
\begin{aligned}
Q_{Oil_{t}}&=\alpha(\gamma)+\sum_{i=1}^{p}\phi_{i}(\gamma)Oil_{t-i}+\sum_{i=0}^{q_{1}}\omega_{i}(\gamma)EPU_{t-i}+\sum_{i=0}^{q_{2}}\lambda_{i}(\gamma)S\&P500_{t-i}+\sum_{i=0}^{q_{3}}\theta_{i}(\gamma)CSI300_{t-i}
\\
&+\sum_{i=0}^{q_{4}}\psi_{i}(\gamma)Interest_{t-i}+\sum_{i=0}^{q_{5}}\delta_{i}(\gamma)Panic_{t-i}+\varepsilon_{t}(\gamma)
\end{aligned}
\end{equation}

Eq.~(\ref{eq:2}) is the basic form of QARDL$(p,q)$ model, where $\gamma$ is the quantile index relates to \{0.1, 0.2, 0.3, 0.4, 0.5, 0.6, 0.7, 0.8 and 0.9\}. $\varepsilon_{t}(\gamma)=Oil_{t}- Q_{Oil_{t}}(\gamma/F_{t-1})$ where $Q_{Oil_{t}}(\gamma/F_{t-1})$ refers to $\gamma$-th conditional quantile of $Q_{Oil_{t}}$ \citep{kim2003estimation}. To avoid the potential serial correlation problem, we rewrite Eq.~(\ref{eq:2}) into the following form:
\begin{equation}\label{eq:3}
\begin{aligned}
Q_{\Delta{Oil_{t}}}&=\alpha+\rho{Oil_{t-1}}+\phi_{EPU}EPU_{t-1}+\phi_{S\&P500}S\&P500_{t-1}+\phi_{CSI300}CSI300_{t-1}+\phi_{Interest}Interest_{t-1}\\
&+\phi_{Panic}Panic_{t-1}+\sum_{i=1}^{p-1}\phi_{i}\Delta{Oil_{t-i}}+\sum_{i=0}^{q_{1}-1}\omega_{i}\Delta{EPU_{t-i}}+\sum_{i=0}^{q_{2}-1}\lambda_{i}\Delta{S\&P500_{t-i}}+\sum_{i=0}^{q_{3}-1}\theta_{i}\Delta{CSI300_{t-i}}\\
&+\sum_{i=0}^{q_{4}-1}\psi_{i}\Delta{Interest_{t-i}}+\sum_{i=0}^{q_{5}-1}\delta_{i}\Delta{Panic_{t-i}}+\nu_{\gamma}
\end{aligned}
\end{equation}
where $\Delta$ refers to the first difference operator. In Eq.~(\ref{eq:3}), $\Delta{EPU_{t}}$, $\Delta{S\&P500_{t}}$, $\Delta{CSI300_{t}}$, $\Delta{Interest_{t}}$ and $\Delta{Panic_{t}}$ might have a contemporaneously correlation with $\nu_{\gamma}$. To overcome this issue, we employ the following projection of $\nu_{\gamma}$ on $\Delta{EPU_{t}}$, $\Delta{S\&P500_{t}}$, $\Delta{CSI300_{t}}$, $\Delta{Interest_{t}}$ and $\Delta{Panic_{t}}$:
\begin{equation}\notag
\begin{aligned}
\nu_{\gamma}&=\gamma_{EPU}\Delta{EPU_{t}}+\gamma_{S\&P500}\Delta{S\&P500_{t}}+\gamma_{CSI300}\Delta{CSI300_{t}}+\gamma_{Interest}\Delta{Interest_{t}}+\gamma_{Panic}\Delta{Panic_{t}}+\varepsilon_{t}
\end{aligned}
\end{equation}
Now, $\varepsilon_{t}$ is uncorrelated with explanatory variables. Then, we apply this projection into Eq.~(\ref{eq:3}) and obtain the following QARDL-ECM model:
\begin{equation}\label{eq:4}
\begin{aligned}
Q_{\Delta{Oil_{t}}}&=\alpha_{\gamma}+\rho_{\gamma}({Oil_{t-1}}+\beta_{EPU}(\gamma)EPU_{t-1}+\beta_{S\&P500}(\gamma)S\&P500_{t-1}+\beta_{CSI300}(\gamma)CSI300_{t-1}\\
&+\beta_{Interest}(\gamma)Interest_{t-1}+\beta_{Panic}(\gamma)Panic_{t-1})+\sum_{i=1}^{p-1}\phi_{i}\Delta{Oil_{t-i}}+\sum_{i=0}^{q_{1}-1}\omega_{i}\Delta{EPU_{t-i}}\\
&+\sum_{i=0}^{q_{2}-1}\lambda_{i}\Delta{S\&P500_{t-i}}+\sum_{i=0}^{q_{3}-1}\theta_{i}\Delta{CSI300_{t-i}}+\sum_{i=0}^{q_{4}-1}\psi_{i}\Delta{Interest_{t-i}}+\sum_{i=0}^{q_{5}-1}\delta_{i}\Delta{Panic_{t-i}}+\varepsilon_{t}(\gamma)
\end{aligned}
\end{equation}\par	
The long-run parameters are defined as: 
$\beta_{S\&P500}=-\frac{\phi_{S\&P500}}{\rho}$, $\beta_{CSI300}=-\frac{\phi_{CSI300}}{\rho}$, $\beta_{EPU}=-\frac{\phi_{EPU}}{\rho}$, 
$\beta_{Interest}=-\frac{\phi_{Interest}}{\rho}$, 
$\beta_{Panic}=-\frac{\phi_{Panic}}{\rho}$. By applying the delta method, the cumulative short-run parameters are defined as: $\phi_{*}=\sum_{i=1}^{p-1}\phi_{i}$,
$\omega_{*}=\sum_{i=0}^{q_{1}-1}\omega_{i}$,
$\lambda_{*}=\sum_{i=0}^{q_{2}-1}\lambda_{i}$,
$\theta_{*}=\sum_{i=0}^{q_{3}-1}\theta_{i}$,
$\psi_{*}=\sum_{i=0}^{q_{4}-1}\psi_{i}$,
$\delta_{*}=\sum_{i=0}^{q_{5}-1}\delta_{i}$. The ECM parameter is presented as $\rho_{*}$. The estimated value of $\rho_{*}$ should be negative and significant.

\begin{table}[!htp]\addtolength{\tabcolsep}{2pt}
\footnotesize
\caption{Descriptive statistics of daily logarithmic level series.}
\label{Table: ststistics }
\medskip
\centering
\begin{threeparttable}
\begin{tabular}{llllllll}
\hline\hline
\textit{}   &Minimum & Maximum & Mean & Std. Dev & Skewness & Kurtosis & Jarque-Bera   \\
\hline
\textit{Panel A: US}&&&&&&&\\
WTI& 2.2039& 4.1954&3.7556& 0.3352& -1.2664& 2.2342& 152.0261*** \\
USEPU& 4.0337 & 6.6040&5.3833& 0.5543& -0.1385 & -0.6899& 6.9764**\\
S\&P500& 7.7131& 8.3506&8.1339& 0.1354& -0.5350& -0.0800&15.1719***\\
CSI300& 8.1691& 8.6669&8.4358& 0.1245& -0.3534 & -1.0902& 21.8000***\\
Tbill& -0.6539& 0.6098& -0.0514 & 0.3668 & 0.3851& -1.3020& 29.6462***\\
Panic& -4.6052& 2.4230& 1.0287& 0.6199& -4.0816& 32.9603& 15292.9583***\\
& & &&&&&\\
\textit{Panel B: China}&&&&&&&\\
SC& 5.4935& 6.2275 & 5.8183& 0.1681& 0.3345 & -0.9193& 17.0280***\\
CEPU& 3.4809& 6.1040  & 4.8158& 0.4797& -0.1100& -0.0576& 0.6741\\
S\&P500& 7.7131& 8.3506 & 8.1346& 0.1332 & -0.5498& 0.0380& 16.3667***    \\
CSI300& 8.1691& 8.6669& 8.4347& 0.1233& -0.3209& -1.0909& 21.1308***    \\
SHIBOR & 0.3954& 1.1613& 0.7549 & 0.1122& -0.4614& 1.8765& 60.1344***    \\
Panic& -4.6052& 3.1041& 1.5333& 0.9571& -3.0674& 13.2414& 2889.3626***  \\
\hline\hline
\end{tabular}
\begin{tablenotes}[para,flushleft]
Notes:  \textsuperscript{***}, \textsuperscript{**} and \textsuperscript{*} indicate  significance at 1\%, 5\% and 10\% levels, respectively. \\
\end{tablenotes}
\end{threeparttable}
\end{table}

\section{Result}
\label{Sec: result}

Table~\ref{Table: ststistics } illustrates the descriptive statistics of daily logarithmic level series. The mean of WTI futures prices is greater than mean of SC futures prices. The standard deviation of WTI futures is 0.3352 while that statistics of SC futures is 0.1681. These results indicate that volatility for WTI futures market is higher than SC market. Tbill and SC prices are left skewed while other data series are right skewed. The kurtosis statistics of two coronavirus panic series are greater than 3. The results of Jarque-Bera test depict that all of  the variables except CEPU deviated from normal distribution. These statistics results qualify QARDL model is an appropriate method for further analysis.

The ARDL and QARDL model are preferable when variables are integrated of either $I(0)$ or $I(1)$. Therefore, we employ Augmented Dickey-Fuller (ADF) test and Phillips Perron (PP)  test to confirm none of variables is integrated of \textit{I}(2). The results of unit root test are presented in Table~\ref{Table: unit root}. The findings  show that all variables are stationary at the first difference. Hence, we can proceed our analysis. 
		
\begin{table}[!htp]\addtolength{\tabcolsep}{2pt}
\footnotesize
\caption{Results of unit root test.}
\label{Table: unit root}
\medskip
\centering
\begin{threeparttable}
\begin{tabular}{lllll}
\hline\hline
\textit{}& ADF(level)& ADF($\Delta$)& PP(level)& PP($\Delta$)\\
\hline
\multicolumn{2}{l}{\textit{Panel A: US}}&&&\\
WTI & -2.9957 & -5.5761*** & -13.1414 & -329.9850*** \\
USEPU& -3.2244* & -8.7544*** & -64.4737***& -401.5527*** \\
S\&P500 & -3.2079*& -5.9811*** & -16.4971& -454.1951*** \\
CSI300& -2.1788& -7.2543*** & -12.7325 & -305.6564*** \\
Tbill& -3.0488 & -8.1289*** & -8.5860 & -274.7621*** \\
Panic& -2.8220& -9.4032*** & -71.4335***  & -412.4296*** \\
&&&&\\
\multicolumn{2}{l}{\textit{Panel B: China}}&&&\\
SC & -2.8159& -6.8910***& -7.7497& -299.8715*** \\
CEPU& -6.4799***& -11.1268***& -262.3458***& -335.1428*** \\
S\&P500 & 2.7251& -6.0310***& -13.3521& -464.9241*** \\
CSI300& -2.5621 & -7.3213***& -12.1749& -311.1095*** \\
SHIBOR& -3.1304*& -8.6182***& -36.1086***& -264.3322*** \\
Panic & -7.3652***& -8.6817***& -49.9165***& -393.1931*** \\
\hline\hline
\end{tabular}
\begin{tablenotes}[para,flushleft]
Notes:  ADF is the Augmented Dickey-Fuller test of unit root, PP is Phillips Perron test of unit root. The null hypothesis for the ADF and PP is that series is non-stationary. \textsuperscript{***}, \textsuperscript{**} and \textsuperscript{*} indicate  significance at 1\%, 5\% and 10\% levels, respectively. \\
\end{tablenotes}
\end{threeparttable}
\end{table}

\begin{figure}[!htp]
	\includegraphics[scale=0.39]{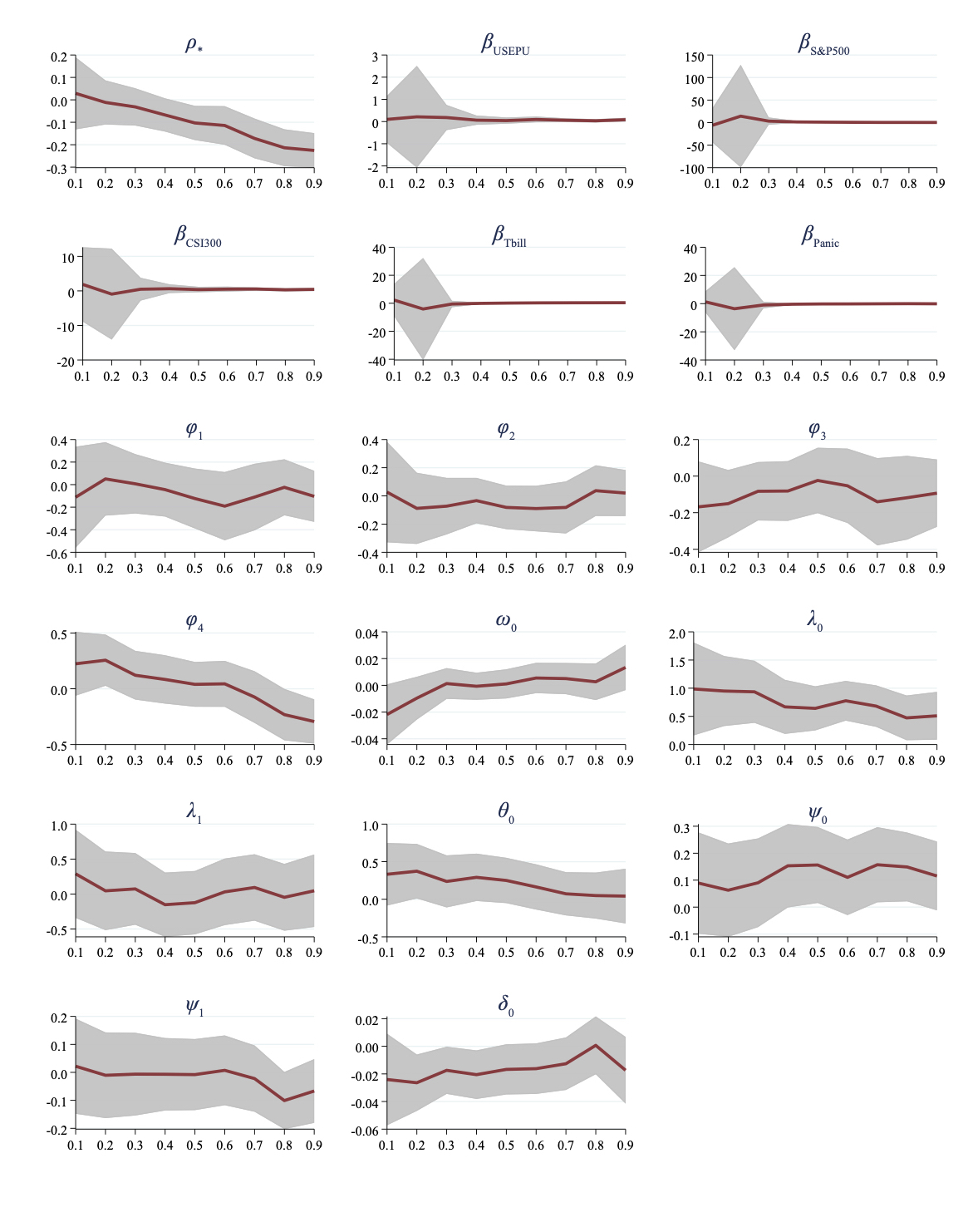}
	\caption{95\% confidence intervals of the QARDL model parameters (US). Horizontal axis indicates the quantile levels (0.1, 0.2, ..., 0.8, 0.9). Vertical axis indicates coefficient estimates of parameters.}
		\label{fig: US}
\end{figure}
\begin{figure}[!htp]
	\centering
	\advance\leftskip-1cm
	\includegraphics[scale=0.35]{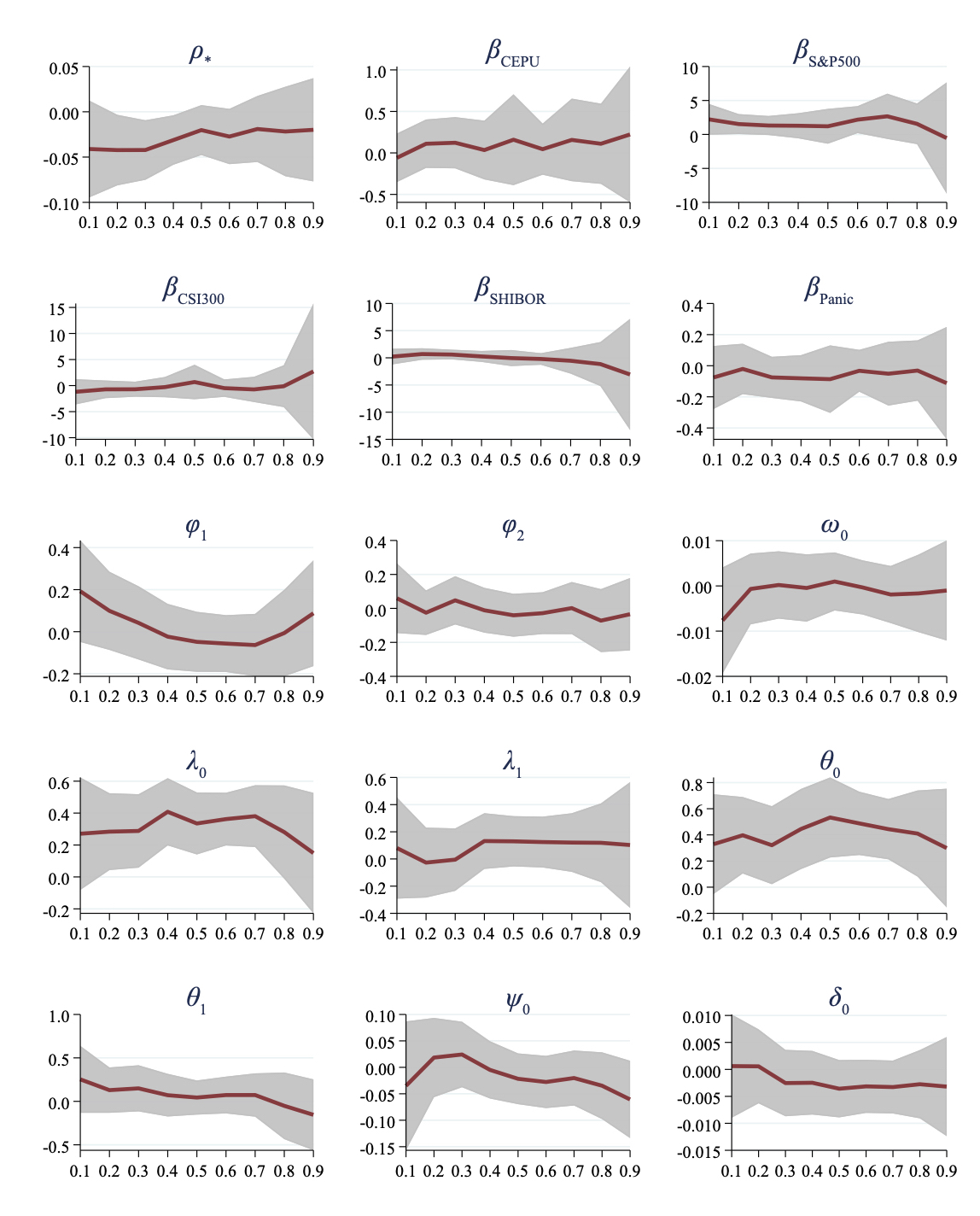}
	\caption{95\% confidence intervals of the QARDL model parameters (China). Horizontal axis indicates the quantile levels (0.1, 0.2, ..., 0.8, 0.9). Vertical axis indicates coefficient estimates of parameters.} 
		\label{fig: China}
\end{figure}

The results of linear ARDL and QARDL model estimation are presented in Table~\ref{Table: US} and Table~\ref{Table: China}.  We report the error correction parameters ({$\rho_{*}$}), the long-run  cointegrating coefficients ({$\beta$}) and short-run coefficients ({$\phi_{i}$},  {$\omega_{i}$}, {$\lambda_{i}$}, {$\theta_{i}$}, {$\psi_{i}$}, {$\delta_{i}$}), along with their standard errors in brackets. Fig.~\ref{fig: US} and Fig.~\ref{fig: China} show the dynamic trends of estimated parameters with a 95\% confidence interval across various quantiles. The figures indicate that most of estimated parameters perform differently at each quanitles.
	 
The ECM parameter measures how strongly crude oil futures prices react to the deviations from the long-run equilibrium relationships. The findings of linear ARDL model show that the ECM coefficient is -0.1082 for US panel and -0.0330 for China panel which means the adjustment speed is 10.82\% and 3.3\%, respectively. Therefore, the adjustment speed of SC futures is slower. For US panel, S\&P500 and Tbill are positively correlated with WTI futures prices in long-run. Furthermore, it can be observed that {$\beta_{Panic}$} and {$\delta_{0}$} are significant. These results indicate an inverse correlation between WTI futures prices and Panic index both in long run and short run. In terms of China panel, the short-run parameters of S\&P500 and CSI300 are significant. More specifically, a 1\% increase in S\&P500 would cause SC futures prices increasing by 23.83\%.  For every 1\% increase in CSI300, SC futures prices would increase by 46.73\%. These results indicate that domestic stock market has a stronger effect on SC futures prices than US stock market. However, the linear ARDL model cannot show the asymmetric relationships between variables. Therefore, we employ the quantile ARDL model.
	 
The results of QARDL estimation explicate that {$\rho_{*}$} is significant and negative in five out of nine quantiles for WTI futures and four out of nine quantiles for SC futures. The empirical findings also indicate that {$\phi_{4}$} is positive at 20\% quantile for US panel. It becomes negative at higher quanitles. In other words, the past changes in WTI futures prices exhibit a heterogeneous and opposite influence on current price levels at extreme quantiles. For China panel, the previous variations in crude oil prices have no effect on current price levels. 

Turning to the results of EPU parameters, we observe that {$\beta_{\mathit{\scriptstyle_{USEPU}}}$} is positive and significant at the highest quantile. It indicates an upward trending long-run relationship between extreme WTI futures price and USEPU. The short-run parameter \textbf{$\omega_{0}$} is negative at 10\%  quantile and positive at 90\% quantile. Our results are partly consistent with \citet{QIN2020117315}. They demonstrate that the effects of USEPU on WTI prices are both negative and positive. Sudden events or policies would change people's sentiments and drive them to change their investment strategies. For China panel, both {$\beta_{\mathit{\scriptstyle_{CEPU}}}$} and \textbf{$\omega_{0}$} are statistically insignificant across all quantiles. It reveals that during our sample period, the changes in CEPU have no impact on SC futures prices. Our results are not aligned with prior research by \citet{10.3389/fenvs.2021.636903}. They use \citet{10.1093/qje/qjw024}'s monthly CEPU index and report that CEPU has a significantly negative impact on the long-term volatility of SC futures.

In term of stock markets, for US panel, the empirical findings show that the short-run parameter \textbf{$\lambda_{0}$} is significant at all quantiles. However, in the long-run, S\&P 500 has no impact on WTI futures prices. Furthermore, S\&P500 plays an important role in SC markets both in short-run and long-run. \textbf{$\lambda_{0}$} is positive and significant across seven quantiles. 
{$\beta_{\mathit{\scriptstyle_{SP500}}}$} is significant at quantiles 10\% to 30\% and 60\%. These findings highlight the strong correlation between US stock market and crude oil markets. The linkage between S\&P500 and crude oil markets has been widely observed and confirmed in the previous studies \citep{KYRTSOU2016239,ZHANG2019192}. In the case of  CSI300, for US panel, {$\beta_{\mathit{\scriptstyle_{CSI300}}}$} is positive and significant at 70\% quantile and 90\% quantile. \textbf{$\theta_{0}$} is significant at low and medium quantiles and indicates a long-run equilibrium between CSI300 and WTI futures prices. CSI300 is also positively correlated with SC prices in short-run. However, the long-run parameter of CSI300 is insignificant. In summary, our results confirm volatility spillovers from stock markets to crude oil markets. In the short-run, crude oil prices are more sensitive to the changes in domestic stock market. However, in the long-run, domestic stock market have limited influence. The price movements of WTI futures are driven by CSI300. Similarly, SC futures prices are driven by variations in S\&P500. 

Our results also indicate that current changes in Tbill are positively correlated with  WTI futures prices while past changes in Tbill have a negative impact on it. The co-movements between interest rates and crude oil prices are sensitive to abnormal global and macro-economic events \citep{MENSI2020100836}. The nexus between US interest rates and WTI prices have been reported in previous literature \citep{ARORA2013546}.
Interest rates might change levels of inventory and market participants' expectations \citep{WANG2013792,MENSI2020100836}. Consequently, interest rates would affect the supply and demand fundamentals and drive oil prices.
For China panel, SHIBOR is not correlated with SC prices in any case. Due to the outbreak of COVID-19 pandemic, global oil prices dropped sharply on 2020. At that time, Chinese companies imported a large volume of crude oil. Therefore, variations in interest rates have no substantial effect on inventory levels of China.

Furthermore, the results exhibit that coronavirus panic has a short-term negative impact on WTI futures prices. Our findings are consistent with results reported by \citet{ATRI2021102061}. COVID-19 pandemic has a profound influence on supply and demand of crude oil. In 2020, global aggregate demand for oil dropped significantly due to travel restrictions. The crude oil producers of US reduced drilling activity in response to rapid decline in oil demand. Therefore, there is a negative linkage between Panic index and WTI futures prices. Interestingly, $\beta_{Panic}$ and $\delta_{0}$ are insignificant in China panel. It means that COVID-19 pandemic has limited influence on SC futures prices. These results do not support \citet{NIU2021102173} who find coronavirus panic has significant influence on volatility of SC prices. Our results are supported by the fact that China's total oil imports increased by 7.3\% in 2020. As the largest oil-importing country, the sharp decline in global oil demand has a relatively small impact on China. Therefore, Panic index cannot drive SC futures prices. In summary, our results demonstrate that oil-importing country and oil-exporting might respond differently to COVID-19 pandemic shocks.

\begin{landscape}
	\begin{table}\centering
		\setlength\tabcolsep{2pt}		
		\scriptsize 
		\caption{Linear ARDL and QARDL estimation results for US.}
		\label{Table: US}
		\medskip
		\centering
		\begin{threeparttable}
			\begin{tabular}{llllllllllllllllllll}
				\hline\hline
				Linear ARDL&\textbf{$\alpha$}&\textbf{$\rho_{*}$}&\textbf{$\beta_{\mathit{\scriptstyle_{USEPU}}}$}&\textbf{$\beta_{\scriptstyle_{S\&P500}}$}&\textbf{$\beta_{\scriptstyle_{CSI300}}$}&\textbf{$\beta_{\mathit{\scriptstyle_{Tbill}}}$}&\textbf{$\beta_{\mathit{\scriptstyle_{Panic}}}$}&\textbf{$\phi_{1}$}&\textbf{$\phi_{2}$}&\textbf{$\phi_{3}$}&\textbf{$\phi_{4}$}&\textbf{$\omega_{0}$}&\textbf{ $\lambda_{0}$}&\textbf{$\lambda_{1}$} & \textbf{$\theta_{0}$} &\textbf{$\psi_{0}$}&\textbf{$\psi_{1}$}& \textbf{$\delta_{0}$}\\
				\hline
				& -1.0316*** & -0.1082*** & 0.0029& 0.0515& 0.1249  & 0.0098 & -0.0467*** & -0.0669& -0.0843 & -0.1915***& 0.1553*** & -0.0062 & 0.9738***& -0.2939& 0.2012& 0.1295**& -0.0017  & -0.0315**\\
				&(0.3969)& (0.0263)& (0.0151)& (0.0949)& (0.0814)	& (0.0176) 	& (0.0158)	& (0.0583)& (0.0532) & (0.0550)	& (0.0549)	& (0.0127) 	& (0.2240)	& (0.2348)& (0.2648)& (0.0658)	& (0.0653)& (0.0156) \\
				&&&&&&&&&&&&&&&&&&\\
				QARDL&	\textbf{$\alpha$}    &\textbf{$\rho_{*}$}       & \textbf{$\beta_{\mathit{\scriptstyle_{USEPU}}}$} &\textbf{$\beta_{\scriptstyle_{S\&P500}}$} & \textbf{$\beta_{\scriptstyle_{CSI300}}$} &\textbf {$\beta_{\mathit{\scriptstyle_{Tbill}}}$} & \textbf{$\beta_{\mathit{\scriptstyle_{Panic}}}$} & \textbf{$\phi_{1}$}     & \textbf{$\phi_{2}$}    & \textbf{$\phi_{3}$}  & \textbf{$\phi_{4}$}   &\textbf{$\omega_{0}$}   &\textbf{ $\lambda_{0}$}   &\textbf{$\lambda_{1}$} & \textbf{$\theta_{0}$}      &\textbf{$\psi_{0}$}  &\textbf{$\psi_{1}$}   & \textbf{$\delta_{0}$}  \\
				\hline
				0.1 & -0.9896& 0.0304& 0.1038& -5.3707& 1.7290& 2.1646&1.1672 & -0.1173& 0.0231& -0.1725  & 0.2186& -0.0219* & 0.9913** & 0.2951& 0.3407& 0.0847& 0.0209& -0.0232\\
				& (0.9017)& (0.0867)& (0.5450)& (19.5559) & (5.4697)& (5.7165)  & (3.6249)& (0.2582) & (0.1733)& (0.1359) & (0.1383) & (0.0115)& (0.4190)& (0.3326) & (0.2840)& (0.0986) & (0.0869)& (0.0178)\\
				0.2 & -1.2197**& -0.0114  & 0.2196& 14.5194 & -0.8741& -4.1766& -3.6587 & 0.0504 & -0.0878 & -0.1506  & 0.2563**   & -0.0095 & 0.9505***& 0.0534   & 0.3735* & 0.0627   & -0.0106 & -0.0264**\\
				& (0.5684)& (0.0523) & (1.2591)  & (61.6178) & (7.0361) & (20.0538)  & (16.1686)  & (0.1847)& (0.1143) & (0.0981) & (0.1173)   & (0.0079) & (0.3102) & (0.2780) & (0.1982)  & (0.0887) & (0.0728) & (0.0106) \\
				0.3 & -0.8706**& -0.0314  & 0.1687   & 3.3048 & 0.5060  & -0.5357 & -0.9663& 0.0070 & -0.0719  & -0.0825  & 0.1214  & 0.0011 & 0.9340*** & 0.0753   & 0.2365 & 0.0916   & -0.0070 & -0.0173* \\
				& (0.4280) & (0.0439 & (0.3186) & (4.1220) & (1.7316)   & (1.1910)  & (1.2182)& (0.1514)  & (0.0922) & (0.0862) & (0.1193)   & (0.0065) & (0.2904) & (0.2387) & (0.1634) & (0.0849) & (0.0727)& (0.0094)\\
				0.4 & -0.9479** & -0.0672 & 0.0646 & 1.5516 & 0.6253  & -0.0787  & -0.4167*& -0.0457& -0.0331& -0.0821  & 0.0822 & -0.0008 & 0.6741*** & -0.1560  & 0.2999* & 0.1534** & -0.0046 & -0.0204** \\
				& (0.3830) & (0.0412) & (0.1161) & (1.0685)  & (0.7383) & (0.2740) & (0.2252) & (0.1288)& (0.0820)& (0.0871) & (0.1126)   & (0.0060) & (0.2433) & (0.2218) & (0.1604)  & (0.0773) & (0.0659)& (0.0094)\\
				0.5 & -0.8602**& -0.1027** & 0.0440& 1.0933& 0.3887& 0.1446 & -0.2188** & -0.1235 & -0.0805 & -0.0233  & 0.0396& 0.0012 & 0.6420*** & -0.1223  & 0.2491* & 0.1565** & -0.0080  & -0.0166* \\
				& (0.4151) & (0.0458) & (0.0786) & (0.6741) & (0.4411) & (0.1402)& (0.1011) & (0.1369) & (0.0807)& (0.0944) & (0.1141) & (0.0056)  & (0.2084) & (0.2257) & (0.1455) & (0.0749) & (0.0689)& (0.0098)\\
				0.6 & -0.7957*& -0.1142**& 0.0952 & 0.7590 & 0.5139 & 0.291** & -0.1808**& -0.1909& -0.0899& -0.0525  & 0.0437 & 0.0055 & 0.7760*** & 0.0325 & 0.1632 & 0.1111 & 0.0073 & -0.0163  \\
				& (0.4183) & (0.0491)  & (0.0730) & (0.5927) & (0.3627) & (0.1190)& (0.0869) & (0.1393)& (0.0801)& (0.1032) & (0.1192)   & (0.0057) & (0.1817) & (0.2361) & (0.1354) & (0.0730) & (0.0644) & (0.0099) \\
				0.7 & -0.8395*& -0.1716***& 0.0600 & 0.4940& 0.5403** & 0.3319*** & -0.1041 & -0.1092  & -0.0810 & -0.1408  & -0.0741 & 0.0051& 0.6801***& 0.0931   & 0.0726 & 0.1580** & -0.0229 & -0.0126 \\
				& (0.4393) & (0.0497)& (0.0497)& (0.3762) & (0.2335)& (0.0932)& (0.0645)  & (0.1342) & (0.0899) & (0.1064) & (0.1311) & (0.0066) & (0.2018)& (0.2341) & (0.1269) & (0.0710) & (0.0567)& (0.0106) \\
				0.8 & -0.6640 & -0.2121*** & 0.0357 & 0.5127  & 0.3234 & 0.3733*** & -0.0228 & -0.0206 & 0.0383 & -0.1195  & -0.2307* & 0.0028 & 0.4729** & -0.0418  & 0.0465 & 0.1500** & -0.0998* & 0.0010 \\
				& (0.4303)  & (0.0445) & (0.0437)& (0.3208) & (0.2005)& (0.0854) & (0.0521) & (0.1196)& (0.0844) & (0.0950) & (0.1226) & (0.0077) & (0.2131) & (0.2500) & (0.1337) & (0.0703) & (0.0567)& (0.0108) \\
				0.9 & -0.7632* & -0.2255***& 0.0849* & 0.4057 & 0.4392**& 0.3918*** & -0.1028**& -0.1038& 0.0207& -0.0931  & -0.2931*** & 0.0134*& 0.5107**& 0.0480 & 0.0415& 0.1155   & -0.0669 & -0.0173\\
				& (0.4532)  & (0.0398) & (0.0434)& (0.3318) & (0.2140) & (0.0846) & (0.0495) & (0.1024) & (0.0818)& (0.0900) & (0.0971) & (0.0079) & (0.2172) & (0.2471) & (0.1647)& (0.0721) & (0.0581)  & (0.0113)   \\
				\hline\hline
			\end{tabular}
			\begin{tablenotes}[para,flushleft]
				Notes: The standard errors are in parentheses. \textsuperscript{***}, \textsuperscript{**} and \textsuperscript{*} indicate  significance at 1\%, 5\% and 10\% levels, respectively. 
			\end{tablenotes}
		\end{threeparttable}
	\end{table}
\end{landscape}

\begin{landscape}
	\begin{table}\centering
		\setlength\tabcolsep{2pt}
		\footnotesize
		\caption{Linear ARDL and QARDL estimation results for China.}
		\label{Table: China}
		\medskip
		\centering
		\begin{threeparttable}
			\begin{tabular}{llllllllllllllllll}
				\hline\hline
				\hline
				Linear ARDL&	\textbf{$\alpha$} &\textbf{$\rho_{*}$} & \textbf{$\beta_{\mathit{\scriptstyle_{CEPU}}}$} &\textbf{$\beta_{\scriptstyle_{S\&P500}}$} & \textbf{$\beta_{\scriptstyle_{CSI300}}$} &\textbf {$\beta_{\mathit{\scriptstyle_{SHIBOR}}}$} & \textbf{$\beta_{\mathit{\scriptstyle_{Panic}}}$} & \textbf{$\phi_{1}$}     & \textbf{$\phi_{2}$}     &\textbf{$\omega_{0}$} &\textbf{ $\lambda_{0}$} &\textbf{$\lambda_{1}$} & \textbf{$\theta_{0}$}  &\textbf{$\theta_{1}$}  &\textbf{$\psi_{0}$}     & \textbf{$\delta_{0}$}  \\
				\hline
				&-0.1025 &-0.0330*** & 0.0010& 0.0337& 0.0028& -0.0082 & -0.0020 & 0.0843& 0.0334& -0.0023 & 0.2383*** & 0.0442 & 0.4673***& 0.0226& -0.0283  & -0.0009 \\
				& (0.1090)& (0.0123) & (0.0037)& (0.0287) & (0.0269) 	& (0.0145)& (0.0019)& (0.0573) & (0.0535)	& (0.0028)& (0.0768)& (0.0786)& (0.0986) 	& (0.0971)	& (0.0248)	& (0.0025) \\
				&&&&&&&&&&&&&&&&\\
				QARDL&	\textbf{$\alpha$}    &\textbf{$\rho_{*}$}       & \textbf{$\beta_{\mathit{\scriptstyle_{CEPU}}}$} &\textbf{$\beta_{\scriptstyle_{S\&P500}}$} & \textbf{$\beta_{\scriptstyle_{CSI300}}$} &\textbf {$\beta_{\mathit{\scriptstyle_{SHIBOR}}}$} & \textbf{$\beta_{\mathit{\scriptstyle_{Panic}}}$} & \textbf{$\phi_{1}$}     & \textbf{$\phi_{2}$}     &\textbf{$\omega_{0}$}   &\textbf{ $\lambda_{0}$}   &\textbf{$\lambda_{1}$} & \textbf{$\theta_{0}$}    &\textbf{$\theta_{1}$}  &\textbf{$\psi_{0}$}     & \textbf{$\delta_{0}$}  \\
				\hline
				0.1 & -0.1093  & -0.0411& -0.0590 & 2.222*& -1.1925& 0.2322 & -0.0755 & 0.1926 & 0.0598 & -0.0076  & 0.2706 & 0.0805 & 0.3292* & 0.2532 & -0.0350  & 0.0006   \\
				& (0.2497) & (0.0273)  & (0.1494)  & (1.1532) & (-1.2320)& (0.7335) & (0.1034)& (0.1232) & (0.1048) & (0.0060) & (0.1798)  & (0.1901) & (0.1946)  & (0.1966) & (0.0622) & (0.0049) \\
				0.2 & -0.0835 & -0.0423** & 0.1107  & 1.5280**  & -0.7209 & 0.7007 & -0.0205 & 0.0993 & -0.0251 & -0.0007 & 0.2836** & -0.0267 & 0.3972*** & 0.1290 & 0.0185 & 0.0006 \\
				& (0.1774) & (0.0199)  & (0.1489)  & (0.7491)& (0.8641) & (0.5409) & (0.0827) & (0.0949) & (0.0669) & (0.0040) & (0.1228)  & (0.1314) & (0.1494)  & (0.1333) & (0.0382) & (0.0035) \\
				0.3 & -0.0082  & -0.0422** & 0.1225 & 1.3163* & -0.6998 & 0.6060 & -0.0748 & 0.0420 & 0.0474 & 0.0002 & 0.2882** & -0.0049  & 0.3212**  & 0.1497 & 0.0243 & -0.0025  \\
				& (0.1815) & (0.0168)  & (0.1576)  & (0.7170) & (0.7377) & (0.4552) & (0.0675) & (0.0889) & (0.0726) & (0.0038) & (0.1174)  & (0.1173) & (0.1522)  & (0.1355) & (0.0316) & (0.0031) \\
				0.4 & -0.0789  & -0.0312** & 0.0347 & 1.2800 & -0.2925& 0.2581& -0.0809 & -0.0235  & -0.0110  & -0.0005  & 0.4080*** & 0.1318& 0.4449*** & 0.0711& -0.0048 & -0.0025  \\
				& (0.1624) & (0.0139) & (0.1808)  & (0.9436)& (0.9932)& (0.5149) & (90.0760) & (0.0795) & (0.0674) & (0.0038) & (0.1075)  & (0.1047) & (0.1565)  & (0.1253) & (0.0276) & (0.0030) \\
				0.5 & -0.2095 & -0.0202 & 0.1590 & 1.2000 & 0.6850& -0.0420 & -0.0863 & -0.0481 & -0.0404 & 0.0010 & 0.3350*** & 0.1301& 0.5331*** & 0.0438 & -0.0217  & -0.0036  \\
				& (0.1271) & (0.0141) & (0.2793) & (1.3088) & (1.6873)  & (0.7563) & (0.1107) & (0.0728) & (0.0645) & (0.0033) & (0.0991)  & (0.0944) & (0.1562)  & (0.1003) & (0.0245) & (0.0027) \\
				0.6 & -0.2087* & -0.0273* & 0.0452 & 2.1777** & -0.4902 & -0.2077 & -0.0324 & -0.0564 & -0.0280 & -0.0003 & 0.3625*** & 0.1246 & 0.4881*** & 0.0728 & -0.0277 & -0.0031  \\
				& (0.1244) & (0.0156) & (0.1572)  & (1.0136) & (0.8535) & (0.5433) & (0.0687) & (0.0690) & (0.0629) & (0.0031) & (0.0842)  & (0.0954) & (0.1236)  & (0.1079) & (0.0252) & (0.0025) \\
				0.7 & -0.1823& -0.0190 & 0.1570 & 2.6722 & -0.7363 & -0.5373 & -0.0511& -0.0635  & 0.0018 & -0.0019  & 0.3807*** & 0.1209 & 0.4438*** & 0.0725   & -0.0202  & -0.0033  \\
				& (0.1311) & (0.0186) & (0.2539)  & (1.6998) & (1.2355) & (1.2106) & (0.1046) & (0.0757) & (0.0783) & (0.0032) & (0.0989)  & (0.1102) & (0.1178)& (0.1273) & (0.0265) & (0.0025) \\
				0.8 & -0.1067  & -0.0217 & 0.1103  & 1.5666 & -0.1232 & -1.1455  & -0.0310 & -0.0062 & -0.0718 & -0.0017  & 0.2808* & 0.1191& 0.4101**  & -0.0516 & -0.0347  & -0.0027  \\
				& (0.1792) & (0.0252)  & (0.2466) & (1.5271) & (2.0476) & (2.0661) & (0.0990) & (0.1051) & (0.0946) & (0.0044) & (0.1492)  & (0.1479) & (0.1682)& (0.1959) & (0.0322) & (0.0032) \\
				0.9 & -0.1999  & -0.0199 & 0.2227 & -0.5307 & 2.7174 & -3.0586  & -0.1120 & 0.0879 & -0.0342  & -0.0010 & 0.1495 & 0.1028 & 0.2990 & -0.1551  & -0.0607  & -0.0032  \\
				& (0.2442) & (0.0291) & (0.4177) & (4.1810) & (6.6517) & (5.2209) & (0.1843) & (0.1284) & (0.1088) & (0.0056) & (0.1926)  & (0.2362) & (0.2319)  & (0.2083) & (0.0373) & (0.0047)\\
				\hline\hline
			\end{tabular}
			\begin{tablenotes}[para,flushleft]
				Notes: The standard errors are in parentheses. \textsuperscript{***}, \textsuperscript{**} and \textsuperscript{*} indicate  significance at 1\%, 5\% and 10\% levels, respectively. 
			\end{tablenotes}
		\end{threeparttable}
	\end{table}
\end{landscape}

\section{Conclusion}
\label{Sec: conclusion}
This study employs the ARDL model and QARDL model to explore the impact of economic policy uncertainty, stock markets, interest rates and coronavirus panic on crude oil markets during COVID-19 pandemic period. This paper finds some interesting results of these price drivers. First, compared with SC futures market, WTI futures market is more sensitive to the fluctuations in economic policy uncertainty. USEPU has a significant impact on extreme WTI futures prices while CEPU has no influence on SC futures prices. Second, our results support the presence of spillover effects between stock markets and crude oil markets. More specifically, in the short-run, crude oil market is positively and strongly influenced by domestic stock market. In the long-run, WTI futures prices are driven by variations in CSI300. Similarly, the movements of S\&P500 plays a crucial role in determining changes in SC futures prices. Third, the results indicate that WTI futures prices are driven by US interest rates both in long-run and short-run. However, SHIBOR does not have impact on SC futures prices. Fourth, we observe a significant negative linkage between coronavirus panic and WTI futures prices. However, panic is insignificant for China panel over all quantiles. 

In conclusion, the determinants of WTI futures and SC futures prices are quiet different. There exist several possible explanations for the results. First, the US is an oil-exporting country while China is an oil-importing country. Therefore, the direct (such as oil production) and indirect (such as interest rates) effects of supply and demand shocks are different in US and China. Second, WTI futures and SC futures have diverse properties, including different trading regulations and investment environment. As a new-developed market, SC has limited pricing power and is less efficient in market operation mechanism \citep{ZHANG2021120050}. Third, our explanatory variables are chosen based on determinants of mature crude oil markets. These factors might have limited effects on the newly launched market during our sample period. For these reasons, some factors play prominent roles in movements of WTI futures prices but have no influence on SC market. Further study need to be done to examine other factors' 
impact on SC futures, especially the effects of supply and demand fundamentals.

Our study provides practical and significant implications for investors and policymakers. WTI futures market is driven by economic policy uncertainty, stock markets, interest rates and coronavirus panic. Therefore, portfolio managers and individual investors should pay attention to fluctuations in stock markets to manage risk efficiently. When adjusting monetary policies, regulators should concern financial stability and transmission mechanism. According to World Health Organization, impacts of COVID-19 pandemic would be long-lasting. To maintain the stability of crude oil futures market, US government should restore investors' sentiment and oversee risks from financial markets.

Although Chinese stock market has a strong positive effect on Shanghai crude oil futures in the short-run,  the market participants also need to concern risks and volatility of US stock market. SC futures still has a long way to become an Asian benchmark. China government should improve the investment environment and trading mechanism of SC futures market. Further research is required to better understand the price mechanism of SC futures as well as its relationship with international oil benchmarks.

	
	

\begin{thebibliography}{35}
\expandafter\ifx\csname natexlab\endcsname\relax\def\natexlab#1{#1}\fi
\providecommand{\url}[1]{\texttt{#1}}
\providecommand{\href}[2]{#2}
\providecommand{\path}[1]{#1}
\providecommand{\DOIprefix}{doi:}
\providecommand{\ArXivprefix}{arXiv:}
\providecommand{\URLprefix}{URL: }
\providecommand{\Pubmedprefix}{pmid:}
\providecommand{\doi}[1]{\href{http://dx.doi.org/#1}{\path{#1}}}
\providecommand{\Pubmed}[1]{\href{pmid:#1}{\path{#1}}}
\providecommand{\bibinfo}[2]{#2}
\ifx\xfnm\relax \def\xfnm[#1]{\unskip,\space#1}\fi
\bibitem[{Aloui et~al.(2016)Aloui, Gupta and Miller}]{ALOUI201692}
\bibinfo{author}{Aloui, R.}, \bibinfo{author}{Gupta, R.},
  \bibinfo{author}{Miller, S.M.}, \bibinfo{year}{2016}.
\newblock \bibinfo{title}{Uncertainty and crude oil returns}.
\newblock \bibinfo{journal}{Energy Economics} \bibinfo{volume}{55},
  \bibinfo{pages}{92--100}.
\newblock \DOIprefix\doi{10.1016/j.eneco.2016.01.012}.
\bibitem[{Apergis et~al.(2021)Apergis, Hayat and Saeed}]{APERGIS2021112118}
\bibinfo{author}{Apergis, N.}, \bibinfo{author}{Hayat, T.},
  \bibinfo{author}{Saeed, T.}, \bibinfo{year}{2021}.
\newblock \bibinfo{title}{Us partisan conflict uncertainty and oil prices}.
\newblock \bibinfo{journal}{Energy Policy} \bibinfo{volume}{150},
  \bibinfo{pages}{112--118}.
\newblock \DOIprefix\doi{10.1016/j.enpol.2020.112118}.
\bibitem[{Arora and Tanner(2013)}]{ARORA2013546}
\bibinfo{author}{Arora, V.}, \bibinfo{author}{Tanner, M.},
  \bibinfo{year}{2013}.
\newblock \bibinfo{title}{Do oil prices respond to real interest rates?}
\newblock \bibinfo{journal}{Energy Economics} \bibinfo{volume}{36},
  \bibinfo{pages}{546--555}.
\newblock \DOIprefix\doi{10.1016/j.eneco.2012.11.001}.
\bibitem[{Atri et~al.(2021)Atri, Kouki and imen Gallali}]{ATRI2021102061}
\bibinfo{author}{Atri, H.}, \bibinfo{author}{Kouki, S.}, \bibinfo{author}{imen
  Gallali, M.}, \bibinfo{year}{2021}.
\newblock \bibinfo{title}{The impact of covid-19 news, panic and media coverage
  on the oil and gold prices: An ardl approach}.
\newblock \bibinfo{journal}{Resources Policy} \bibinfo{volume}{72},
  \bibinfo{pages}{102061}.
\newblock \DOIprefix\doi{https://doi.org/10.1016/j.resourpol.2021.102061}.
\bibitem[{Baker et~al.(2016)Baker, Bloom and Davis}]{10.1093/qje/qjw024}
\bibinfo{author}{Baker, S.R.}, \bibinfo{author}{Bloom, N.},
  \bibinfo{author}{Davis, S.J.}, \bibinfo{year}{2016}.
\newblock \bibinfo{title}{{Measuring Economic Policy Uncertainty}}.
\newblock \bibinfo{journal}{The Quarterly Journal of Economics}
  \bibinfo{volume}{131}, \bibinfo{pages}{1593--1636}.
\newblock \DOIprefix\doi{10.1093/qje/qjw024}.
\bibitem[{Baumeister and Peersman(2013)}]{baumeister2013time}
\bibinfo{author}{Baumeister, C.}, \bibinfo{author}{Peersman, G.},
  \bibinfo{year}{2013}.
\newblock \bibinfo{title}{Time-varying effects of oil supply shocks on the us
  economy}.
\newblock \bibinfo{journal}{American Economic Journal: Macroeconomics}
  \bibinfo{volume}{5}, \bibinfo{pages}{1--28}.
\newblock \DOIprefix\doi{10.1257/mac.5.4.1}.
\bibitem[{Benkraiem et~al.(2018)Benkraiem, Lahiani, Miloudi and
  Shahbaz}]{BENKRAIEM2018169}
\bibinfo{author}{Benkraiem, R.}, \bibinfo{author}{Lahiani, A.},
  \bibinfo{author}{Miloudi, A.}, \bibinfo{author}{Shahbaz, M.},
  \bibinfo{year}{2018}.
\newblock \bibinfo{title}{New insights into the us stock market reactions to
  energy price shocks}.
\newblock \bibinfo{journal}{Journal of International Financial Markets,
  Institutions and Money} \bibinfo{volume}{56}, \bibinfo{pages}{169--187}.
\newblock \DOIprefix\doi{10.1016/j.intfin.2018.02.004}.
\bibitem[{Chen et~al.(2016)Chen, Liao, Tang and Wei}]{CHEN201642}
\bibinfo{author}{Chen, H.}, \bibinfo{author}{Liao, H.}, \bibinfo{author}{Tang,
  B.J.}, \bibinfo{author}{Wei, Y.M.}, \bibinfo{year}{2016}.
\newblock \bibinfo{title}{Impacts of opec's political risk on the international
  crude oil prices: An empirical analysis based on the svar models}.
\newblock \bibinfo{journal}{Energy Economics} \bibinfo{volume}{57},
  \bibinfo{pages}{42--49}.
\newblock \DOIprefix\doi{10.1016/j.eneco.2016.04.018}.
\bibitem[{Cho et~al.(2015)Cho, hwan Kim and Shin}]{CHO2015281}
\bibinfo{author}{Cho, J.S.}, \bibinfo{author}{hwan Kim, T.},
  \bibinfo{author}{Shin, Y.}, \bibinfo{year}{2015}.
\newblock \bibinfo{title}{Quantile cointegration in the autoregressive
  distributed-lag modeling framework}.
\newblock \bibinfo{journal}{Journal of Econometrics} \bibinfo{volume}{188},
  \bibinfo{pages}{281--300}.
\newblock \DOIprefix\doi{10.1016/j.jeconom.2015.05.003}.
\bibitem[{Guo et~al.(2021)Guo, Li, Li and You}]{GUO2021105198}
\bibinfo{author}{Guo, Y.}, \bibinfo{author}{Li, J.}, \bibinfo{author}{Li, Y.},
  \bibinfo{author}{You, W.}, \bibinfo{year}{2021}.
\newblock \bibinfo{title}{The roles of political risk and crude oil in stock
  market based on quantile cointegration approach: A comparative study in china
  and us}.
\newblock \bibinfo{journal}{Energy Economics} \bibinfo{volume}{97},
  \bibinfo{pages}{105198}.
\newblock \DOIprefix\doi{10.1016/j.eneco.2021.105198}.
\bibitem[{Huang and Luk(2020)}]{huang2020measuring}
\bibinfo{author}{Huang, Y.}, \bibinfo{author}{Luk, P.}, \bibinfo{year}{2020}.
\newblock \bibinfo{title}{Measuring economic policy uncertainty in china}.
\newblock \bibinfo{journal}{China Economic Review} \bibinfo{volume}{59},
  \bibinfo{pages}{101367}.
\newblock \DOIprefix\doi{10.1016/j.chieco.2019.101367}.
\bibitem[{Joo et~al.(2020)Joo, Suh, Lee and Ahn}]{JOO2020100516}
\bibinfo{author}{Joo, K.}, \bibinfo{author}{Suh, J.H.}, \bibinfo{author}{Lee,
  D.}, \bibinfo{author}{Ahn, K.}, \bibinfo{year}{2020}.
\newblock \bibinfo{title}{Impact of the global financial crisis on the crude
  oil market}.
\newblock \bibinfo{journal}{Energy Strategy Reviews} \bibinfo{volume}{30},
  \bibinfo{pages}{100516}.
\newblock \DOIprefix\doi{10.1016/j.esr.2020.100516}.
\bibitem[{Kilian(2009)}]{kilian2009not}
\bibinfo{author}{Kilian, L.}, \bibinfo{year}{2009}.
\newblock \bibinfo{title}{Not all oil price shocks are alike: Disentangling
  demand and supply shocks in the crude oil market}.
\newblock \bibinfo{journal}{American Economic Review} \bibinfo{volume}{99},
  \bibinfo{pages}{1053--1069}.
\newblock \DOIprefix\doi{10.1257/aer.99.3.1053}.
\bibitem[{Kim and White(2003)}]{kim2003estimation}
\bibinfo{author}{Kim, T.H.}, \bibinfo{author}{White, H.}, \bibinfo{year}{2003}.
\newblock \bibinfo{title}{Estimation, inference, and specification testing for
  possibly misspecified quantile regression}.
\newblock \bibinfo{publisher}{Emerald Group Publishing Limited}.
\newblock \DOIprefix\doi{10.1016/S0731-9053(03)17005-3}.
\bibitem[{Kyrtsou et~al.(2016)Kyrtsou, Mikropoulou and Papana}]{KYRTSOU2016239}
\bibinfo{author}{Kyrtsou, C.}, \bibinfo{author}{Mikropoulou, C.},
  \bibinfo{author}{Papana, A.}, \bibinfo{year}{2016}.
\newblock \bibinfo{title}{Does the s\&p500 index lead the crude oil dynamics? a
  complexity-based approach}.
\newblock \bibinfo{journal}{Energy Economics} \bibinfo{volume}{56},
  \bibinfo{pages}{239--246}.
\newblock \DOIprefix\doi{10.1016/j.eneco.2016.02.001}.
\bibitem[{Lei et~al.(2019)Lei, Shang, Chen and Wei}]{LEI2019341}
\bibinfo{author}{Lei, L.}, \bibinfo{author}{Shang, Y.}, \bibinfo{author}{Chen,
  Y.}, \bibinfo{author}{Wei, Y.}, \bibinfo{year}{2019}.
\newblock \bibinfo{title}{Does the financial crisis change the economic risk
  perception of crude oil traders? a midas quantile regression approach}.
\newblock \bibinfo{journal}{Finance Research Letters} \bibinfo{volume}{30},
  \bibinfo{pages}{341--351}.
\newblock \DOIprefix\doi{10.1016/j.frl.2018.10.016}.
\bibitem[{Lu et~al.(2020)Lu, Li, Chai and Wang}]{LU2020104721}
\bibinfo{author}{Lu, Q.}, \bibinfo{author}{Li, Y.}, \bibinfo{author}{Chai, J.},
  \bibinfo{author}{Wang, S.}, \bibinfo{year}{2020}.
\newblock \bibinfo{title}{Crude oil price analysis and forecasting: A
  perspective of “new triangle”}.
\newblock \bibinfo{journal}{Energy Economics} \bibinfo{volume}{87},
  \bibinfo{pages}{104721}.
\newblock \DOIprefix\doi{10.1016/j.eneco.2020.104721}.
\bibitem[{Meng and Liu(2019)}]{MENG2019476}
\bibinfo{author}{Meng, F.}, \bibinfo{author}{Liu, L.}, \bibinfo{year}{2019}.
\newblock \bibinfo{title}{Analyzing the economic sources of oil price
  volatility: An out-of-sample perspective}.
\newblock \bibinfo{journal}{Energy} \bibinfo{volume}{177},
  \bibinfo{pages}{476--486}.
\newblock \DOIprefix\doi{10.1016/j.energy.2019.04.161}.
\bibitem[{Mensi et~al.(2020a)Mensi, Rehman and Al-Yahyaee}]{MENSI2020100836}
\bibinfo{author}{Mensi, W.}, \bibinfo{author}{Rehman, M.U.},
  \bibinfo{author}{Al-Yahyaee, K.H.}, \bibinfo{year}{2020}a.
\newblock \bibinfo{title}{Time-frequency co-movements between oil prices and
  interest rates: Evidence from a wavelet-based approach}.
\newblock \bibinfo{journal}{The North American Journal of Economics and
  Finance} \bibinfo{volume}{51}, \bibinfo{pages}{100836}.
\newblock \DOIprefix\doi{10.1016/j.najef.2018.08.019}.
\bibitem[{Mensi et~al.(2020b)Mensi, Sensoy, Vo and Kang}]{MENSI2020101829}
\bibinfo{author}{Mensi, W.}, \bibinfo{author}{Sensoy, A.}, \bibinfo{author}{Vo,
  X.V.}, \bibinfo{author}{Kang, S.H.}, \bibinfo{year}{2020}b.
\newblock \bibinfo{title}{Impact of covid-19 outbreak on asymmetric
  multifractality of gold and oil prices}.
\newblock \bibinfo{journal}{Resources Policy} \bibinfo{volume}{69},
  \bibinfo{pages}{101829}.
\newblock \DOIprefix\doi{10.1016/j.resourpol.2020.101829}.
\bibitem[{Niu et~al.(2021)Niu, Liu, Gao and Zhang}]{NIU2021102173}
\bibinfo{author}{Niu, Z.}, \bibinfo{author}{Liu, Y.}, \bibinfo{author}{Gao,
  W.}, \bibinfo{author}{Zhang, H.}, \bibinfo{year}{2021}.
\newblock \bibinfo{title}{The role of coronavirus news in the volatility
  forecasting of crude oil futures markets: Evidence from china}.
\newblock \bibinfo{journal}{Resources Policy} \bibinfo{volume}{73},
  \bibinfo{pages}{102173}.
\newblock \DOIprefix\doi{10.1016/j.resourpol.2021.102173}.
\bibitem[{Pan et~al.(2017)Pan, Wang, Wu and Yin}]{PAN2017130}
\bibinfo{author}{Pan, Z.}, \bibinfo{author}{Wang, Y.}, \bibinfo{author}{Wu,
  C.}, \bibinfo{author}{Yin, L.}, \bibinfo{year}{2017}.
\newblock \bibinfo{title}{Oil price volatility and macroeconomic fundamentals:
  A regime switching garch-midas model}.
\newblock \bibinfo{journal}{Journal of Empirical Finance} \bibinfo{volume}{43},
  \bibinfo{pages}{130--142}.
\newblock \DOIprefix\doi{10.1016/j.jempfin.2017.06.005}.
\bibitem[{Pesaran et~al.(1999)Pesaran, Shin and Smith}]{pesaran1999pooled}
\bibinfo{author}{Pesaran, M.H.}, \bibinfo{author}{Shin, Y.},
  \bibinfo{author}{Smith, R.P.}, \bibinfo{year}{1999}.
\newblock \bibinfo{title}{Pooled mean group estimation of dynamic heterogeneous
  panels}.
\newblock \bibinfo{journal}{Journal of the American statistical Association}
  \bibinfo{volume}{94}, \bibinfo{pages}{621--634}.
\newblock \DOIprefix\doi{10.2307/2670182}.
\bibitem[{Prest(2018)}]{PREST201863}
\bibinfo{author}{Prest, B.C.}, \bibinfo{year}{2018}.
\newblock \bibinfo{title}{Explanations for the 2014 oil price decline: Supply
  or demand?}
\newblock \bibinfo{journal}{Energy Economics} \bibinfo{volume}{74},
  \bibinfo{pages}{63--75}.
\newblock \DOIprefix\doi{10.1016/j.eneco.2018.05.029}.
\bibitem[{Qin et~al.(2020a)Qin, Su, Hao and Tao}]{QIN2020117315}
\bibinfo{author}{Qin, M.}, \bibinfo{author}{Su, C.W.}, \bibinfo{author}{Hao,
  L.N.}, \bibinfo{author}{Tao, R.}, \bibinfo{year}{2020}a.
\newblock \bibinfo{title}{The stability of u.s. economic policy: Does it really
  matter for oil price?}
\newblock \bibinfo{journal}{Energy} \bibinfo{volume}{198},
  \bibinfo{pages}{117315}.
\newblock \DOIprefix\doi{10.1016/j.energy.2020.117315}.
\bibitem[{Qin et~al.(2020b)Qin, Hong, Chen and Zhang}]{QIN2020104851}
\bibinfo{author}{Qin, Y.}, \bibinfo{author}{Hong, K.}, \bibinfo{author}{Chen,
  J.}, \bibinfo{author}{Zhang, Z.}, \bibinfo{year}{2020}b.
\newblock \bibinfo{title}{Asymmetric effects of geopolitical risks on energy
  returns and volatility under different market conditions}.
\newblock \bibinfo{journal}{Energy Economics} \bibinfo{volume}{90},
  \bibinfo{pages}{104851}.
\newblock \DOIprefix\doi{https://doi.org/10.1016/j.eneco.2020.104851}.
\bibitem[{Wang and Chueh(2013)}]{WANG2013792}
\bibinfo{author}{Wang, Y.S.}, \bibinfo{author}{Chueh, Y.L.},
  \bibinfo{year}{2013}.
\newblock \bibinfo{title}{Dynamic transmission effects between the interest
  rate, the us dollar, and gold and crude oil prices}.
\newblock \bibinfo{journal}{Economic Modelling} \bibinfo{volume}{30},
  \bibinfo{pages}{792--798}.
\newblock \DOIprefix\doi{10.1016/j.econmod.2012.09.052}.
\bibitem[{Wei et~al.(2017)Wei, Liu, Lai and Hu}]{WEI2017141}
\bibinfo{author}{Wei, Y.}, \bibinfo{author}{Liu, J.}, \bibinfo{author}{Lai,
  X.}, \bibinfo{author}{Hu, Y.}, \bibinfo{year}{2017}.
\newblock \bibinfo{title}{Which determinant is the most informative in
  forecasting crude oil market volatility: Fundamental, speculation, or
  uncertainty?}
\newblock \bibinfo{journal}{Energy Economics} \bibinfo{volume}{68},
  \bibinfo{pages}{141--150}.
\newblock \DOIprefix\doi{10.1016/j.eneco.2017.09.016}.
\bibitem[{Yang(2019)}]{YANG2019219}
\bibinfo{author}{Yang, L.}, \bibinfo{year}{2019}.
\newblock \bibinfo{title}{Connectedness of economic policy uncertainty and oil
  price shocks in a time domain perspective}.
\newblock \bibinfo{journal}{Energy Economics} \bibinfo{volume}{80},
  \bibinfo{pages}{219--233}.
\newblock \DOIprefix\doi{10.1016/j.eneco.2019.01.006}.
\bibitem[{Yi et~al.(2021)Yi, Yang and Li}]{10.3389/fenvs.2021.636903}
\bibinfo{author}{Yi, A.}, \bibinfo{author}{Yang, M.}, \bibinfo{author}{Li, Y.},
  \bibinfo{year}{2021}.
\newblock \bibinfo{title}{Macroeconomic uncertainty and crude oil futures
  volatility–evidence from china crude oil futures market}.
\newblock \bibinfo{journal}{Frontiers in Environmental Science}
  \bibinfo{volume}{9}, \bibinfo{pages}{21}.
\newblock \DOIprefix\doi{10.3389/fenvs.2021.636903}.
\bibitem[{Zhang et~al.(2021)Zhang, Di and Farnoosh}]{ZHANG2021120050}
\bibinfo{author}{Zhang, Q.}, \bibinfo{author}{Di, P.},
  \bibinfo{author}{Farnoosh, A.}, \bibinfo{year}{2021}.
\newblock \bibinfo{title}{Study on the impacts of shanghai crude oil futures on
  global oil market and oil industry based on vecm and dag models}.
\newblock \bibinfo{journal}{Energy} \bibinfo{volume}{223},
  \bibinfo{pages}{120050}.
\newblock \DOIprefix\doi{10.1016/j.energy.2021.120050}.
\bibitem[{Zhang and Hamori(2021)}]{ZHANG2021101702}
\bibinfo{author}{Zhang, W.}, \bibinfo{author}{Hamori, S.},
  \bibinfo{year}{2021}.
\newblock \bibinfo{title}{Crude oil market and stock markets during the
  covid-19 pandemic: Evidence from the us, japan, and germany}.
\newblock \bibinfo{journal}{International Review of Financial Analysis}
  \bibinfo{volume}{74}, \bibinfo{pages}{101702}.
\newblock \DOIprefix\doi{10.1016/j.irfa.2021.101702}.
\bibitem[{Zhang and Wang(2019)}]{ZHANG2019192}
\bibinfo{author}{Zhang, Y.J.}, \bibinfo{author}{Wang, J.L.},
  \bibinfo{year}{2019}.
\newblock \bibinfo{title}{Do high-frequency stock market data help forecast
  crude oil prices? evidence from the midas models}.
\newblock \bibinfo{journal}{Energy Economics} \bibinfo{volume}{78},
  \bibinfo{pages}{192--201}.
\newblock \DOIprefix\doi{10.1016/j.eneco.2018.11.015}.
\bibitem[{Zhang and Yan(2020)}]{ZHANG2020750}
\bibinfo{author}{Zhang, Y.J.}, \bibinfo{author}{Yan, X.X.},
  \bibinfo{year}{2020}.
\newblock \bibinfo{title}{The impact of us economic policy uncertainty on wti
  crude oil returns in different time and frequency domains}.
\newblock \bibinfo{journal}{International Review of Economics \& Finance}
  \bibinfo{volume}{69}, \bibinfo{pages}{750--768}.
\newblock \DOIprefix\doi{10.1016/j.iref.2020.04.001}.
\bibitem[{Zhu et~al.(2021)Zhu, Tang, Wei and Lu}]{ZHU2021120949}
\bibinfo{author}{Zhu, P.}, \bibinfo{author}{Tang, Y.}, \bibinfo{author}{Wei,
  Y.}, \bibinfo{author}{Lu, T.}, \bibinfo{year}{2021}.
\newblock \bibinfo{title}{Multidimensional risk spillovers among crude oil, the
  us and chinese stock markets: Evidence during the covid-19 epidemic}.
\newblock \bibinfo{journal}{Energy} \bibinfo{volume}{231},
  \bibinfo{pages}{120949}.
\newblock \DOIprefix\doi{10.1016/j.energy.2021.120949}.

\end{thebibliography}

\end{document}